\documentclass[aps,reprint,showpacs,showkeys,a4paper]{revtex4}
\usepackage{amsmath}
\usepackage{slashed}
\usepackage{graphicx}

\begin{document}

\title{The heavy mesons in Nambu--Jona-Lasinio model} 

\author{Xiao-Yu Guo} 

\author{Xiao-Lin Chen} 

\author{Wei-Zhen Deng}\email{dwz@pku.edu.cn}

\affiliation{School of physics and State Key Laboratory of Nuclear
  Physics and Technology, Peking University, Beijing, 100871, China}

\begin{abstract}
  We propose an extended Nambu--Jona-Lasinio model to include the
  heavy mesons with the heavy quark symmetry. The quark
  current-current interaction is generalized to include the heavy
  quark currents. In order to comply with the heavy quark spin
  symmetry at the heavy quark limit, the quark mass dependence of the
  interaction strength is introduced. The light and heavy
  pseudo-scalar and vector meson, their masses and the weak decay
  constants, are calculated in the unified frame.
\end{abstract}



\maketitle

\section{Introduction}

In recent years, some exotic hadron states have been observed in
experiments. Many of them cannot be explained easily as the
conventional quarkonia. A possible interpretation is the hypothesis of
molecular state \cite{Tornqvist:2004qy, Swanson:2003tb,
  Liu:2008tn}. Many of the studies on the exotic states were based on
the heavy quark effective theory (HQET). Recently the chiral quark
model have been used in solving the molecular
state~\cite{Liu:2008qb,Yu:2011wb}. Currently, the most difficulty to
identify a molecular state is the uncertainty of the parameters about
the interaction strengths and the form factors.

On principle these parameters can be calculated from QCD on the quark
level. However, in the low-energy region where the QCD perturbation
method fails we have to rely on effective theories. Among them, the
Nambu--Jona-Lasinio (NJL) model \cite{Nambu:1961tp,Nambu:1961fr} was
widely used to investigate many low-energy hadron problems related to
the QCD symmetries in a simple way
\cite{Klevansky:1992qe,Hatsuda:1994pi,Vogl:1991qt}.

By means of Dyson-Schwinger equation (DSE), the dynamic quark mass is
generated from the spontaneous chiral symmetry breaking. After solving
the Bethe-Salpeter equation (BSE), pseudo-scalar mesons are obtained
as the Goldstone bosons \cite{Nambu:1961tp,Nambu:1961fr}. Other mesons
such as vector mesons and axial-vector mesons were included by
introducing more chiral invariant interactions \cite{Bernard:1988db,
  Blin:1990um, Takizawa:1991mx}. Also the model was extended to
comprise the strange flavor \cite{Bernard:1987sg, Klimt:1989pm}. A
bosonization technique was also developed \cite{Eguchi:1974cg} and
many works were done along this approach \cite{Ebert:1982pk,
  Ebert:1985kz, Reinhardt:1988xu, Bijnens:1995ww}.

Because of the QCD color coulomb interaction, a heavy quark spin
symmetry is reached in the heavy quark limit that the dependency of
hadronic matrix elements on the orientation of the heavy quark spin
vanishes \cite{Isgur:1989vq}. From the heavy quark symmetry the HQET
formalism was developed (for a review, see Refs.~\cite{manoh,
  Neubert:1993mb}).

Some efforts were made on studying the heavy mesons within the NJL
model \cite{Ebert:1994tv,Mota:2006ex}. The bosonization technique was
used in these studies to obtained the meson Lagrangian in HQET. Using
the heavy quark propagators in the heavy quark limit, the DSE+BSE
approach was also been used to calculate heavy meson observables
\cite{Ivanov:1997yg,Ivanov:1998ms}.

In the NJL model study, the color-octet vector current interaction
$(\bar{\psi}\lambda_C^a\gamma_\mu\psi)
(\bar{\psi}\lambda_C^a\gamma^\mu\psi)$ was widely adopted since it is
closely related to the QCD interaction. In many DSE+BSE
calculations such as in Ref.~\cite{Roberts:2000aa}, the interaction
between two quarks was assumed to be intermediated by the gluon with a
complicated effective propagator. So the color-octet vector current of
the quark should be dominant. The DSE+BSE calculation using the gluon
propagator was also performed in the heavy meson case
\cite{Nguyen:2009if}. If we naively treat the gluon propagator as a
constant in the coordinate space, we would obtain an NJL model with
the color-octet vector current interaction. In the heavy quark limit
where the heavy quark mass $m_Q $ tends to infinity, we will show that
the heavy quark spin symmetry is valid only for the color-octet vector
current interaction.

Other contact interactions such as the color-octet axial-vector
current interaction $(\bar{\psi}\lambda_C^a\gamma_\mu\gamma_5\psi)
(\bar{\psi}\lambda_C^a\gamma^\mu\gamma_5\psi)$ are needed to give a
more comprehensive description of the light flavor mesons such as the
$\rho$ meson \cite{Bernard:1988db, Blin:1990um, Takizawa:1991mx,
  Bernard:1987sg, Klimt:1989pm}. We will show that the heavy quark
spin symmetry would not be reached if these interactions exist in the
heavy quark limit. To maintain this symmetry, these interactions
should be considered as higher order terms and should be $1/m_Q$
suppressed. This is critical to extend the NJL model to include heavy
quark flavors.

We will extend the NJL model to include the heavy quark flavors.  The
typical approach of DSE+BSE will be used to obtain properties of heavy
mesons. In this way, we can calculate the mass splitting between the
pseudo-scalar mesons $D$ (or $B$) and the vector mesons $D^*$ (or
$B^*$) which is the effect of finite heavy quark mass according to the
heavy quark expansion.  Due to the fact that the heavy quark masses
are far beyond the NJL cutoff scale, the usual 4-dimensional cutoff is not
appropriate here. We will use the 3-dimensional cutoff following
Refs.~\cite{Nambu:1961tp, Nambu:1961fr, Bernard:1988db}.

In the next section, we will generalize the NJL interaction to include
the heavy quark flavor and derive the mass dependence of the coupling
strength parameters according to the heavy quark spin symmetry. In
Section~\ref{BSEM}, we will give a brief account of the DSE+BSE
formalism to treat the quark and meson states. In
Section~\ref{HeavyQlimit}, we will take the heavy quark limit and
demonstrate the heavy quark spin symmetry. In Section~\ref{NumResult},
numerical calculation will be performed and the result will be
compared to the empirical data. Finally we will give a brief summary.

\section{NJL Interaction with Heavy Quark Symmetry}
\label{QNJL}

In many NJL studies, when dealing with the three light flavors
$q=u,d,s$, the interaction was taken to be the color current
interaction
\begin{equation}
  \mathcal{L}_4= G_V(\bar{q}\lambda_C^a\gamma_\mu q)^2
  +G_A(\bar{q}\lambda_C^a\gamma_\mu\gamma_5 q)^2\label{L_basic}.
\end{equation}
The interaction maintains the $U_f(3)\otimes U_f(3)\otimes SU_C(3)$
symmetry. Here we will not consider the 6-quark interaction
which was used to deal the $U_A(1)$ anomaly, since we will not
concern the anomaly here and the contribution of the anomaly term is
small \cite{Klimt:1989pm}. After a Fierz transformation, we can get a Fierz
invariant interaction
\begin{equation}
  \mathcal{L}_4^F=\frac{4}{9}G_1\sum_{i=0}^8\left[(\bar{q}\lambda_f^iq)^2
    +(\bar{q}i\gamma_5\lambda_f^iq)^2\right]
  -\frac{2}{9}G_2\sum_{i=0}^8\left[(\bar{q}\lambda_f^i\gamma_\mu q)^2
    +(\bar{q}\lambda_f^i\gamma_\mu\gamma_5 q)^2\right]
  +\text{color-octet terms}\label{L_basicF},
\end{equation}
where
\begin{equation}
  G_1=G_V-G_A, \quad G_2=G_V+G_A .
\end{equation}
Here the $\lambda_i$'s are the flavor Gell-Mann matrices with
$\lambda_0\equiv \sqrt{\frac{2}{3}} \openone$. The color-octet terms
do not contribute to the DSE+BSE calculation of the meson.

In Ref.~\cite{Ebert:1994tv}, where the heavy flavors $Q=c,b$ were
considered, only the color-octet vector interaction
$(\bar{q}\lambda_C^a\gamma_\mu q)(\bar{Q}\lambda_C^a\gamma^\mu Q)$ was
considered. In Section~\ref{HeavyQlimit}, we will show that only this
term in Eq.~(\ref{L_basic}) respects the heavy quark spin symmetry in
the heavy quark limit.

The color-octet vector interaction is however not enough to describe
the light flavor mesons such as the vector $\rho$ meson.  We will also
show that the heavy quark spin symmetry would not be reached if other
than the vector interaction exists in the heavy quark limit. To
consistently describe the light sector and the heavy sector of the
meson system, we assume that the NJL interaction is originated in the
color-octet vector current. Other currents appear as higher order
correction in some series expansion and thus should be suppressed by
the $1/m_q$ factor if the expansion is taken with respect to the
constituent quark mass $m_q$. According to this thought, we modify the
NJL interaction Eq.~(\ref{L_basic}) to
\begin{equation}
  \mathcal{L}_4=G_V(\bar{q}\lambda_C^a\gamma_\mu q)
  (\bar{q}'\lambda_C^a\gamma_\mu q')^2
  +\frac{h_1}{m_qm_{q'}}(\bar{q}\lambda_C^a\gamma_\mu q)
  (\bar{q}'\lambda_C^a\gamma_\mu q')
  +\frac{h_2}{m_qm_{q'}}(\bar{q}\lambda_C^a\gamma_\mu\gamma_5 q)
  (\bar{q}'\lambda_C^a\gamma_\mu\gamma_5 q').
\end{equation}
Here we can take the light and heavy quarks into a unified frame
$q,q'=u,d,s,c,b$. $h_1$ and $h_2$ are dimensionless parameters. $m_q$
and $m_{q'}$ are the constituent masses of the quarks involved in the
interaction.

\begin{figure}
  \begin{center}
    \includegraphics[width=8cm]{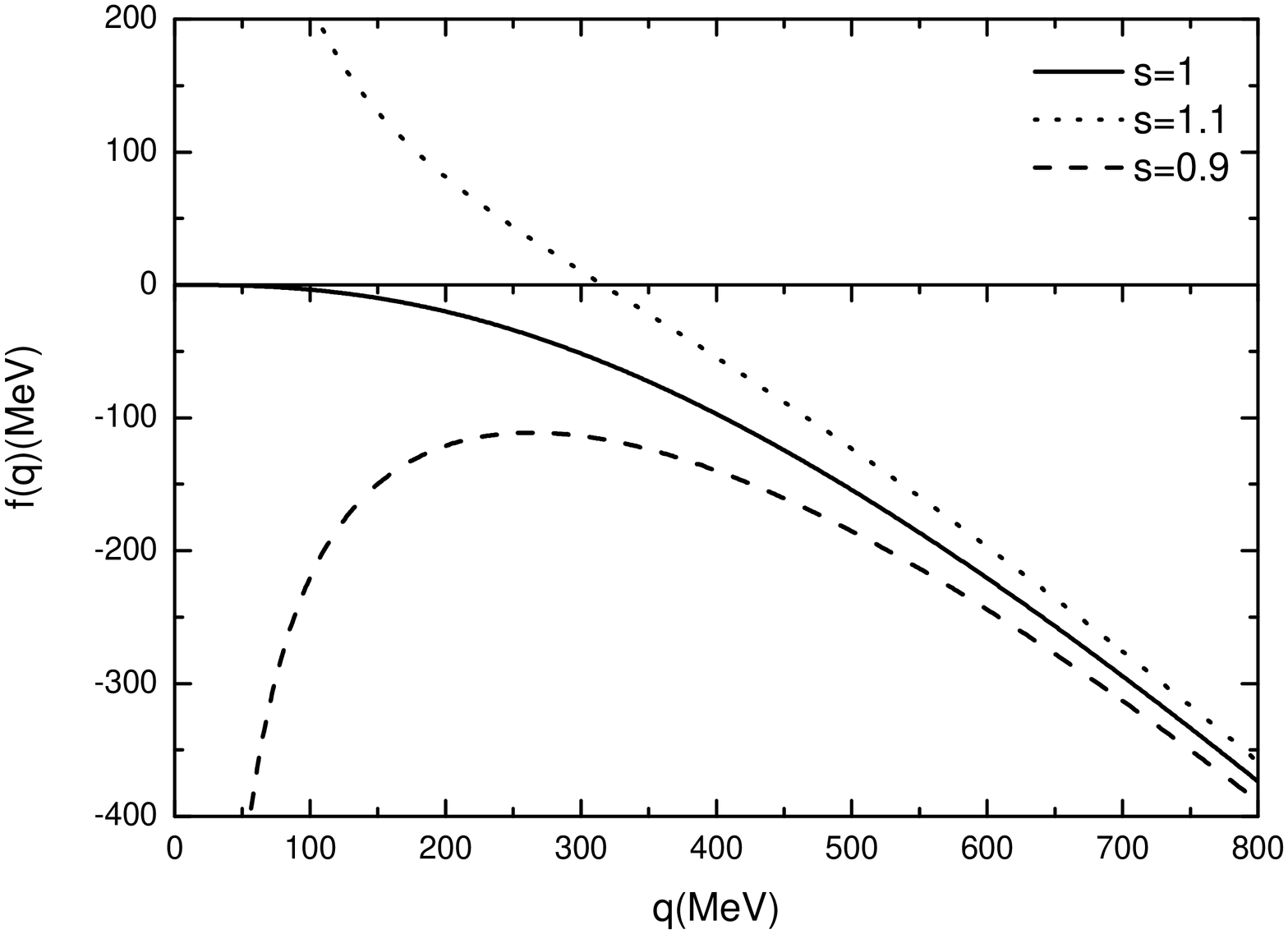}%
    \includegraphics[width=8cm]{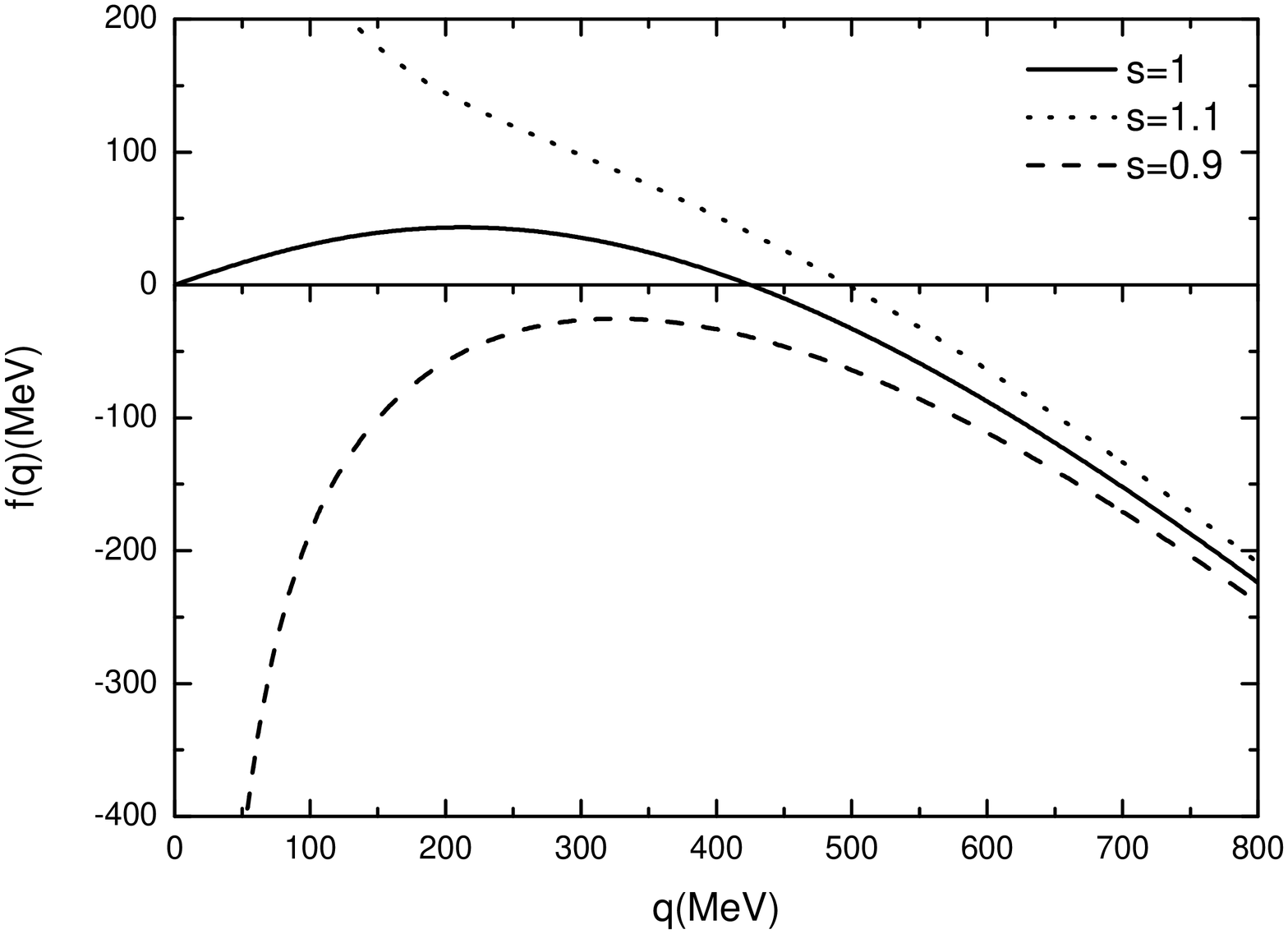}%
    \caption{The gap equation where the zero points are the solution
      of quark mass. The cutoff is taken at
      $\Lambda=750$MeV. $h_2=0.65$, $h_1=sh_2$. The dimensionless
      parameter $g_V$ is defined as $g_V=G_V\Lambda^2$. We show two
      typical situations in the figure: $g_V=gc=9\pi^2/16N_C$ where
      the strength is critically not enough to break the chiral
      symmetry; and $g_V=2.5$ with the chiral symmetry breaking
      solution.}%
    \label{QMassSample}
  \end{center}
\end{figure}

The Fierz invariant interaction Eq.~(\ref{L_basicF}) has the
interaction strengths
\begin{equation}
  G_1=G_V+\frac{h_1-h_2}{m_qm_{q'}}, \quad G_2=G_V+\frac{h_1+h_2}{m_qm_{q'}}.
\end{equation}
We notice that $G_1$ is tightly related to the quark constituent mass
which is the dynamical one generated from the gap equation (see
Eqs.~(\ref{GapEq}) and (\ref{Condensate}) in Section~\ref{BSEM}).  If
$G_1$ depends on the constituent quark mass, the gap equation will
change radically. As shown in FIG.~\ref{QMassSample}, when $h_1=h_2$,
the case of usual NJL model where $G_1=G_V$ is independent on
the constituent quark mass, the gap equation has the two solutions
corresponding to two chiral phases which is believed to exist in
the QCD chiral limit: a Wigner solution at $m_q=0$ and a chiral
symmetry breaking solution at $m_q\neq0$ when the coupling is large
enough.

However, when $h_1\ne h_2$, the gap equation reveals a singularity at $m_q=0$.
Hence, there is no chiral phase (Wigner solution). So we must set
$h_1=h_2=h$ and the NJL interaction turns to be
\begin{equation}
  \mathcal{L}_4=G_V(\bar{\psi}\lambda_C^a\gamma_\mu \psi)^2
  +\frac{h}{m_qm_{q'}}\left[(\bar{\psi}\lambda_C^a\gamma_\mu \psi)^2
    +(\bar{\psi}\lambda_C^a\gamma_\mu\gamma_5 \psi)^2\right]\label{L_optimize}.
\end{equation}
After the Fierz transformation, we obtain the relevant Fierz invariant
interaction: for the light sector
\begin{equation}
  \mathcal{L}_4^F=\frac{4}{9}G_V\left[(\bar{q}\lambda_f^iq)^2
    +(\bar{q}i\gamma_5\lambda_f^iq)^2\right]
  -\frac{2}{9}\left(G_0+\frac{2h}{m_qm_{q'}}\right)
  \left[(\bar{q}\lambda_f^i\gamma_\mu q)^2
    +(\bar{q}\lambda_f^i\gamma_\mu\gamma_5q)^2\right]\label{L_opimize_Fierz1},
\end{equation}
and for the heavy sector
\begin{equation}
  {\mathcal{L}_4^F}'=\frac{8}{9}G_V
  \left[(\bar{Q}q)(\bar{q}Q)+(\bar{Q}i\gamma_5q)(\bar{q}i\gamma_5Q)\right]
  -\frac{4}{9}\left(G_0+\frac{2h}{m_qm_{Q}}\right)
  \left[(\bar{Q}\gamma_\mu q)(\bar{q}\gamma^\mu Q)
    +(\bar{Q}\gamma_\mu\gamma_5q)(\bar{q}\gamma_\mu\gamma_5Q)\right].
\end{equation}

Because the difference between the constituent mass and current mass
of heavy quark should be small, we will ignore the dynamical effect upon
the heavy quark mass in the calculation of heavy mesons.

\section{Bethe-Salpeter Equation and Mesons}
\label{BSEM}
Now we will give a brief account of the method of DSE and BSE used in
our calculation of meson states. Throughout this section we will use
Eq.~(\ref{L_basicF}) as a general form of the NJL interaction.

The Dyson-Schwinger Equation (DSE) is used to obtain the dynamical
quark mass $m_q$. The self-consistent gap equation derived from DSE
reads
\begin{equation}
  m_q=m_{q}^0+\Sigma_q\label{GapEq},
\end{equation}
where $m_q^0$ is the current quark mass and $\Sigma_q$ is the quark
self energy
\begin{equation}
  -i\Sigma_q=i\frac{32}{9}G_1\mathrm{Tr}\int \frac{\mathrm{d}^4p}{(2\pi)^4}
  S_q(p) =-\frac{16G_1}{9}m_qI_1(m_q)\label{Condensate},
\end{equation}
where $S_q(p)$ is the quark propagator. The expression of the integral
$I_1(m_q)$ is given in the Appendix.

We use Bethe-Salpeter equation (BSE) to obtain the meson mass and
amplitude.  The total quark anti-quark scattering amplitude is
obtained from the ladder approximation.  We decompose the amplitude
into different Lorentz structures \cite{Klimt:1989pm}. The relevant
amplitudes are
\begin{align}
  \mathcal{T}_{\mathrm{ps}}=&T_{PP}(i\gamma_5\lambda_i\otimes i\gamma_5\lambda_j)
  +T_{AP}(-i\slashed{\hat{q}}\gamma_5\lambda_i\otimes i\gamma_5\lambda_j)
  +T_{PA}(i\gamma_5\lambda_i\otimes i\slashed{\hat{q}}\gamma_5\lambda_j)
  +T_{AA}^P(-i\slashed{\hat{q}}\gamma_5\lambda_i\otimes
  i\slashed{\hat{q}}\gamma_5\lambda_j), \\
  \mathcal{T}_{\mathrm{v}}=&T_{VV}(\eta^{\mu\nu}\gamma_\mu\lambda_i\otimes
  \gamma_\nu\lambda_j),
\end{align}
where $\hat{q}^\mu=q^\mu/\sqrt{q^2}$,
$\eta_{\mu\nu}=g_{\mu\nu}-\hat{q}^\mu\hat{q}^\nu$. In the ladder approximation,
we need only calculate the loop integral
\[
\includegraphics[height=1.5cm]{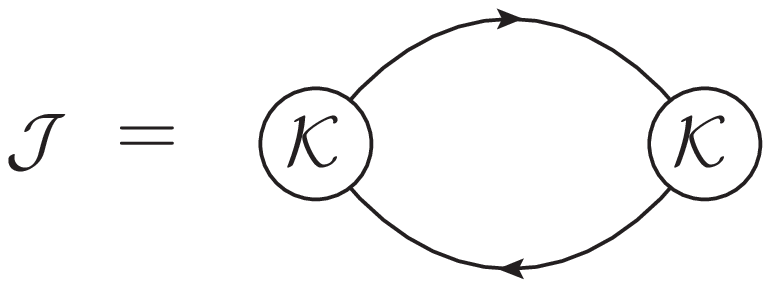}
\]
which can also be decomposed to
\begin{align}
  \mathcal{J}_\mathrm{ps}^{ij}=&
  J_{PP}(i\gamma_5\lambda_i\otimes i\gamma_5\lambda_j)
  +J_{AP}(-i\slashed{\hat{q}}\gamma_5\lambda_i\otimes i\gamma_5\lambda_j)\notag\\
  &+J_{PA}(i\gamma_5\lambda_i\otimes i\slashed{\hat{q}}\gamma_5\lambda_j)
  +J_{AA}^L(-i\slashed{\hat{q}}\gamma_5\lambda_i\otimes
  i\slashed{\hat{q}}\gamma_5\lambda_j),\\
  \mathcal{J}_\mathrm{v}^{ij}=&\eta^{\mu\nu}J_{VV}^T(\gamma_\mu\lambda_i
  \otimes \gamma_\nu\lambda_j).
\end{align}
Then we have
\begin{equation}
  T=\frac{1}{1-J K},
\end{equation}
where
\begin{equation}
  \begin{aligned}
    K_\mathrm{P}&=\frac{16G_1}{9}(i\gamma_5\lambda_i\otimes i\gamma_5\lambda_j),
    & K_\mathrm{S}=&\frac{16G_1}{9}(\lambda_i\otimes \lambda_j)\\
    K_\mathrm{A}&=-\frac{8G_2}{9}(\gamma_\mu\gamma_5\lambda_i
    \otimes \gamma_\nu\gamma_5\lambda_j),&
    K_\mathrm{V}=&-\frac{8G_2}{9}(\gamma_\mu\lambda_i\otimes \gamma_\nu\lambda_j).
  \end{aligned}
\end{equation}
The integrals $J_{AB}$ are defined in ref.~\cite{Klimt:1989pm} and 
formulae with 3-dimensional cut-off are collected in the Appendix.

The meson mass $m_M$ is determined by the pole of the amplitude,
\begin{equation}
  \mathrm{Det}(1-JK)\big|_{q^2=m_M^2}=0 \label{mass-eq} .
\end{equation}

To calculate the weak decay constant of a pseudo-scalar meson,
the quark-meson vertex is obtained by expanding the scattering
amplitude near the meson pole.  For a pseudo-scalar meson $P$, which
could be $\pi$, $K$, $D$, or $B$, the $qqP$ vertex reads
\begin{equation}
  V^i_P(p)=i\gamma_5\lambda^i\left[g_P(p^2)
    -\frac{\slashed{p}}{m_{q}+m_{q'}}\tilde{g}_P(p^2)\right],
\end{equation}
where
\begin{align}
  g^2_P=&\left(\frac{\mathrm{d}D}{\mathrm{d}q^2}\right)_{q^2=m_P^2}^{-1}
  K_P(1-J_{AA}K_A),\\
  \tilde{g}_P=&\frac{m_{q}+m_{q'}}{m_P}\frac{K_AJ_{PA}}{1-J_{AA}K_A}g_P,
\end{align}
where $D=\mathrm{Det}(1-JK)$.  The pion decay constant is given by
\begin{equation}
  \langle0|\bar{q}(0)\gamma^\mu\gamma_5\frac{\lambda_i}{2}q(0)|\pi_j(p)\rangle
  =if_\pi p^\mu\delta_{ij}.
\end{equation}
Similar result holds for the kaon decay constant. In the heavy quark
case the decay constant is given by
\begin{equation}
  \langle0|\bar{q}(0)\gamma^\mu\gamma_5Q(0)|H(p)\rangle=iF_H p^\mu,
\end{equation}
where $H$ could be $D$ or $B$.

\section{Heavy Quark Limit}\label{HeavyQlimit}

In this section, we will discuss the heavy quark limit. After the
Fierz transformation, the relevant interaction between a light quark $q$
and a heavy quark $Q$ in a heavy meson is written in the form
\begin{align}
  {\mathcal{L}_4^F}'=\frac{8}{9}G_1
  \left[(\bar{Q}q)(\bar{q}Q)+(\bar{Q}i\gamma_5q)(\bar{q}i\gamma_5Q)\right]
  -\frac{4}{9}G_2
  \left[(\bar{Q}\gamma_\mu q)(\bar{q}\gamma^\mu Q)
    +(\bar{Q}\gamma_\mu\gamma_5q)(\bar{q}\gamma_\mu\gamma_5Q)\right]
  \label{L_Qq0}.
\end{align}

Consider the heavy meson at rest, $q=m_Hv$, $v=(1,0,0,0)$. In the
heavy quark limit one assumes that the mass difference between $m_H$ of
the heavy meson and $m_Q$ of the heavy quark is a small quantity $l_0$,
\begin{equation}
  m_H=m_Q+l_0\label{m_H}.
\end{equation}
The heavy quark momentum $p$ is expanded around the heavy meson momentum
$q$ as $p=q+k$, where $k$ is assumed to be far smaller than
$m_Q$. Then the propagator of the heavy quark reduces to
\begin{equation}
  \frac{1}{(\slashed{k}+\slashed{q})-m_Q}\approx
  \frac{\slashed{v}+1}{2(k\cdot v+ l_0)} .
\end{equation}
The expression on the right hand side is independent of $m_Q$.  The BSE
loop integrals reduce to
\begin{align}
  J_{PP}&=2iN_C\mathrm{tr}\int\frac{\mathrm{d}^4k}{(2\pi)^4}
  i\gamma_5\frac{1}{\slashed{k}-m_q+i\epsilon}i\gamma_5
  \frac{\slashed{v}+1}{2(k\cdot v+ l_0+i\epsilon)},\\
  J_{PA}&=2iN_C\mathrm{tr}v^\mu\int\frac{\mathrm{d}^4k}{(2\pi)^4}
  i\gamma_5\frac{1}{\slashed{k}-m_q+i\epsilon}(-i\gamma_\mu\gamma_5)
  \frac{\slashed{v}+1}{2(k\cdot v+ l_0+i\epsilon)},\\
  J_{SS}&=2iN_C\mathrm{tr}\int\frac{\mathrm{d}^4k}{(2\pi)^4}
  \frac{1}{\slashed{k}-m_q+i\epsilon}i\gamma_5
  \frac{\slashed{v}+1}{2(k\cdot v+ l_0+i\epsilon)},\\
  J_{SV}&=2iN_C\mathrm{tr}v^\mu\int\frac{\mathrm{d}^4k}{(2\pi)^4}
  \frac{1}{\slashed{k}-m_q+i\epsilon}\gamma_\mu
  \frac{\slashed{v}+1}{2(k\cdot v+ l_0+i\epsilon)},\\
  J_{VV}^{\mu\nu}&=2iN_C\mathrm{tr}\int\frac{\mathrm{d}^4k}{(2\pi)^4}
  \gamma^\mu\frac{1}{\slashed{k}-m_q}\gamma^\nu
  \frac{\slashed{v}+1}{2(k\cdot v+ l_0)},\\
  J_{AA}^{\mu\nu}&=2iN_C\mathrm{tr}\int\frac{\mathrm{d}^4k}{(2\pi)^4}
  \gamma^\mu\gamma_5\frac{1}{\slashed{k}-m_q}\gamma^\nu\gamma_5
  \frac{\slashed{v}+1}{2(k\cdot v+ l_0)}.
\end{align}
We find
\begin{align}
  J_{PP}=J_{PA}&=4iN_C\int\frac{\mathrm{d}^4k}{(2\pi)^4}
  \frac{k\cdot v-m_q}{(k^2-m_q^2+i\epsilon)(v\cdot k +l_0+i\epsilon)},\\
  J_{SS}=J_{SV}&=4iN_C\int\frac{\mathrm{d}^4k}{(2\pi)^4}
  \frac{k\cdot v+m_q}{(k^2-m_q^2+i\epsilon)(v\cdot k +l_0+i\epsilon)},\\
  J_{VV}^{\mu\nu}&=4iN_C\int\frac{\mathrm{d}^4k}{(2\pi)^4}
  \frac{k^\mu v^\nu+v^\mu k^\nu-g^{\mu\nu}k\cdot v+g^{\mu\nu}m_q}
  {(k^2-m_q^2)(v\cdot k +l_0)},\\
  J_{AA}^{\mu\nu}&=4iN_C\int\frac{\mathrm{d}^4k}{(2\pi)^4}
  \frac{k^\mu v^\nu+v^\mu k^\nu-g^{\mu\nu}k\cdot v-g^{\mu\nu}m_q}
  {(k^2-m_q^2)(v\cdot k +l_0)}.
\end{align}
After further decompositions $J_{VV}^{\mu\nu}=J_{VV}^T(g^{\mu\nu}-v^\mu
v^\nu)+J_{VV}^Lv^\mu v^\nu$ and
$J_{AA}^{\mu\nu}=J_{AA}^T(g^{q\mu\nu}-v^\mu v^\nu)+J_{AA}^Lv^\mu
v^\nu$, we have
\begin{align}
  J_{VV}^L&=4iN_C\int\frac{\mathrm{d}^4k}{(2\pi)^4}
  \frac{k\cdot v+m_q}{(k^2-m_q^2)(v\cdot k +l_0)},\\
  J_{VV}^T&=4iN_C\int\frac{\mathrm{d}^4k}{(2\pi)^4}
  \frac{-k\cdot v+m_q}{(k^2-m_q^2)(v\cdot k +l_0)},\\
  J_{AA}^L&=4iN_C\int\frac{\mathrm{d}^4k}{(2\pi)^4}
  \frac{k\cdot v-m_q}{(k^2-m_q^2+i\epsilon)(v\cdot k +l_0+i\epsilon)},\\
  J_{AA}^T&=4iN_C\int\frac{\mathrm{d}^4k}{(2\pi)^4}
  \frac{-k\cdot v-m_q}{(k^2-m_q^2+i\epsilon)(v\cdot k +l_0+i\epsilon)}.
\end{align}
Thus, in the heavy quark limit
\begin{align}
  J_{PP}(l_0)=&J_{PA}(l_0)=J_{AA}^L(l_0)=-J_{VV}^T(l_0),\\
  J_{SS}(l_0)=&J_{SV}(l_0)=J_{VV}^L(l_0)=-J_{AA}^T(l_0).
\end{align}

For a pseudo-scalar meson, the mass equation Eq.~(\ref{mass-eq}) turns
to be
\[
(1-J_{PP}(q^2)K_P)(1-J_{AA}^L(q^2)K_A)-J_{PA}^2(q^2)K_PK_A=0,
\]
which reduces to
\begin{equation}
  1-(K_P+K_A)J_{PP}(l_0)=0,
\end{equation}
in the heavy quark limit.  The mass equation of the
vector partner is $1-J_{VV}^T(q^2)K_V=0$, which leads to
\begin{equation}
  1+K_VJ_{PP}(l_0)=0,
\end{equation}
in the heavy quark limit. If $G_2=G_1$, the mass equations of the
pseudo-scalar meson and the vector meson are identical, and the
heavy quark spin symmetry is obtained.  Otherwise, if $G_2 \ne G_1$,
the mass of the pseudo-scalar meson differs from the mass of the
vector meson.  Similarly, the masses of a scalar meson and its
axial-vector partner will be degenerate in the heavy quark limit if
and only if $G_2=G_1$.

\section{Numerical Results}
\label{NumResult}

\begin{table}
  \caption{Numerical results of the meson masses and decay constants. The
    cal.~I column: results with the NJL interaction
    Eq.~(\ref{L_optimize}).  The cal.~II column: results with the
    interaction Eq.~(\ref{L_basic}) for the light meson sector and the
    interaction Eq.~(\ref{L_Qq0}) for the heavy meson
    sector. The experimental data are taken from
    Ref.~\cite{Nakamura:2010zzi} except for $F_B$ and $F_B^*$ which are taken
    from the lattice calculation in Ref.~\cite{Gamiz:2009ku}
    (see also~\cite{Bazavov:2011aa, Na:2012kp}).}
  \label{TabResultsG0}
  \begin{ruledtabular}
    \begin{tabular}{cccc}
      & cal. I & cal. II & exp. \\
      \hline
      $m_u$(MeV)         &    392  & 389 &\\
      $m_s$(MeV)         &    542  & 540 &\\
      $m_\pi$(MeV)       & 139  & 137  & 135/140\\
      $m_K$(MeV)         & 496  & 496  & 494/498\\
      $f_\pi$(MeV)       & 91.5 & 86.5 & 93.3\\
      $f_K$(MeV)         & 97.9 & 88.6 & 114\\
      $m_\rho$(MeV)      &    771  & 775 & 775\\
      $m_{K^*}$(MeV)     &    918  & 905 & 892\\
      \hline
      $m_D$(GeV)         & 1.87 & 1.86 & 1.86/1.87\\
      $m_{D_s}$(GeV)     & 1.95 & 1.95 & 1.97\\
      $m_{D^*}$(GeV)     & 1.99 & 2.07 & 2.01\\
      $m_{D_s^*}$(GeV)   & 2.12 & 2.20 & 2.11\\
      $m_B$(GeV)         & 5.28 &5.28  & 5.28\\
      $m_{Bs}$(GeV)      & 5.37 &5.37  & 5.37\\
      $m_{B^*}$(GeV)     & 5.31 &5.36  & 5.33\\
      $m_{B_s^*}$(GeV)   & 5.42 &5.47  & 5.42\\
      $F_D$(MeV)         &  139 &123   & 207\\
      $F_{Ds}$(MeV)      &  147 &129   & 258\\
      $F_B$(MeV)         & 96.7 &87.0  & 190 (lattice)\\
      $F_{Bs}$(MeV)      & 107  &91.8  & 231 (lattice)\\
    \end{tabular}
  \end{ruledtabular}
\end{table}

In the NJL interaction Eq.~(\ref{L_optimize}), the input parameters
are the current masses for light quarks and constituent masses for
heavy quarks, the coupling constants and the 3-dimensional cutoff. We used
experimental data of light mesons of $m_\pi$, $m_K$, $m_\rho$, $f_\pi$
to determine parameters $m_{u/d}^0$, $m_s^0$, $G_V$, $h$ and
$\Lambda$. Then the experimental masses of $m_D$ and $m_B$ is used to
determine $m_c$ and $m_b$. The parameters are
\begin{equation}
  \begin{aligned}
    m_{u/d}^0=& 2.79\text{MeV}, & m_s^0=& 72.0\text{MeV}, \\
    m_c=& 1.63\text{GeV}, & m_b=& 4.94\text{GeV}, \\
    \Lambda=& 0.8\text{GeV}, & g_V=&G_V\Lambda^2=2.41, & h=&0.65.
  \end{aligned}
\end{equation}
The resulted masses and weak decay constants are show in the cal.~I
column in TABLE~\ref{TabResultsG0}. We find that the meson mass
spectra, both the light sector and the heavy sector, are well fitted to
the experimental data.  One major difficulty is that the calculated
decay constant decreases with increasing meson mass while the
experimental one increases with mass. As already shown in
Ref.~\cite{Klimt:1989pm}, the theoretical result of $f_K$ is smaller
than the empirical data. In the case of heavy mesons, the theoretical
results are smaller than the empirical ones by almost a factor $2$. We
notice that the decay constant increases with the momentum cutoff
parameter $\Lambda$. A possible explanation is that the momentum
cutoff in heavy sector is larger than in light sector which reflects
the fact that the size of a heavy meson is relatively small.

\begin{figure}
  \begin{center}
    \includegraphics[width=8cm]{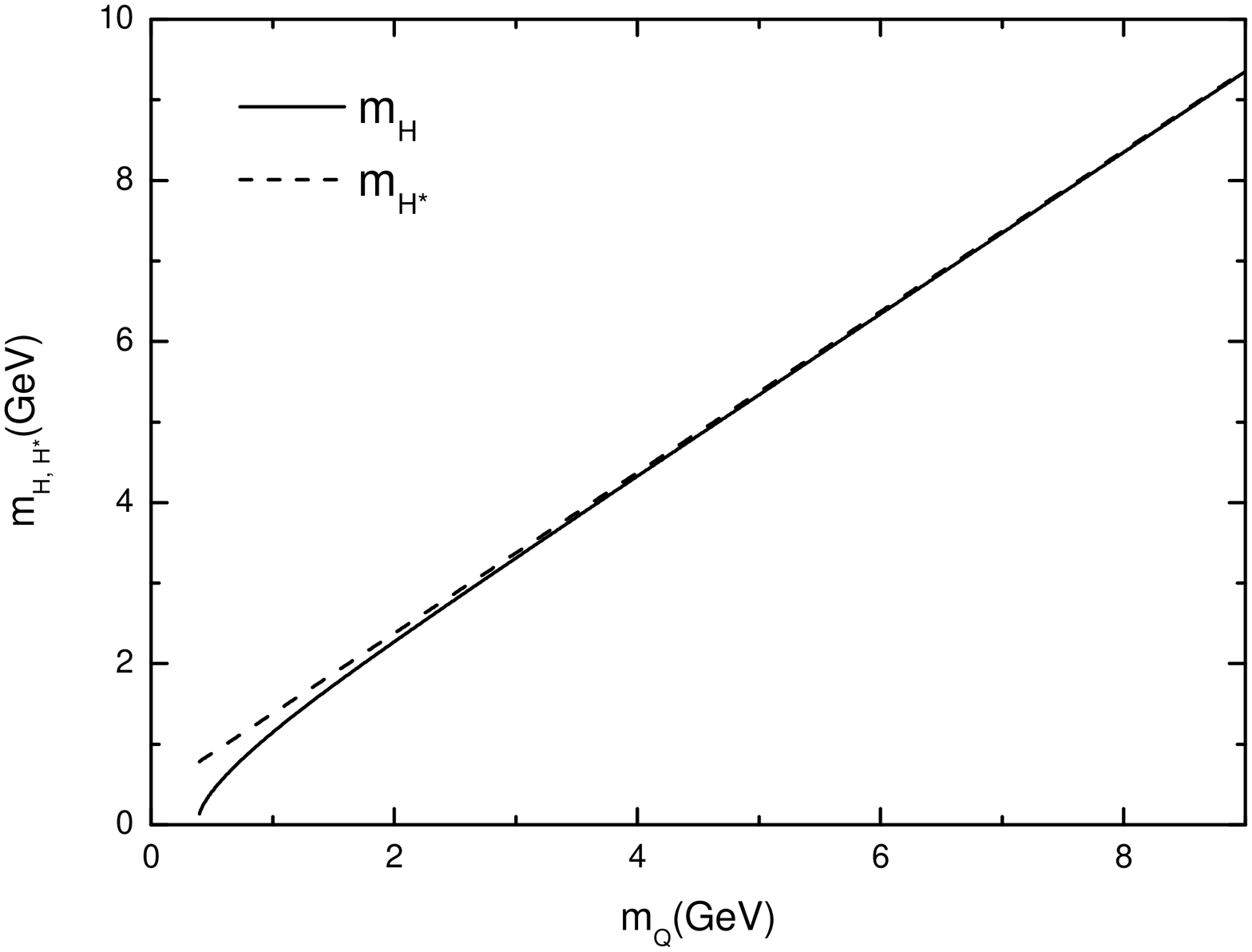}
    \includegraphics[width=8cm]{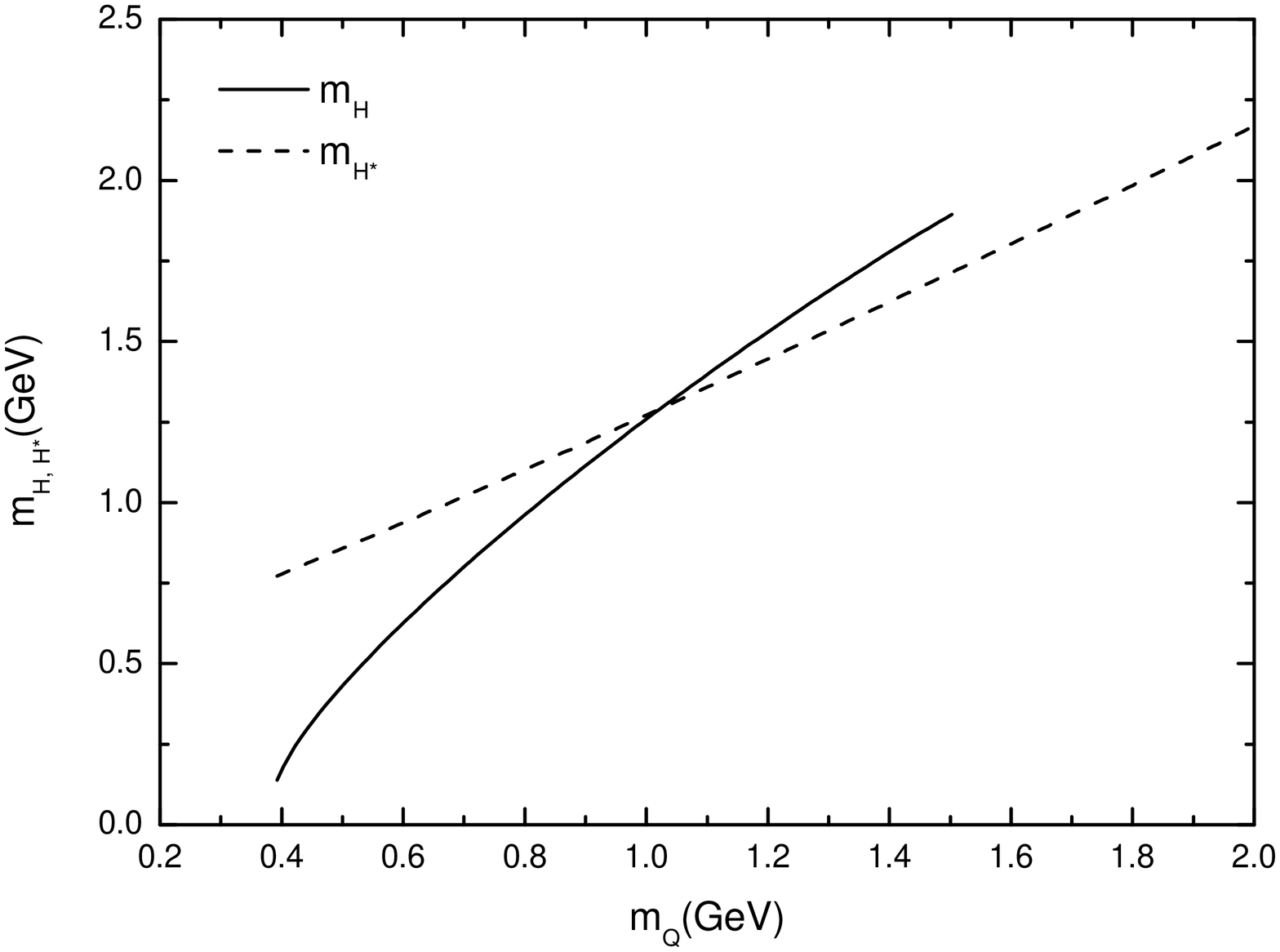}
    \caption{The dependence of heavy-light meson masses on $m_Q$. On
      the left side, with the interaction Eq.~(\ref{L_optimize}). On
      the right side, with the interaction Eq.~(\ref{L_Qq0}). The
      light quark is set to $u$.} \label{DepMcG0Sample1N}
  \end{center}
\end{figure}

The dependence of heavy meson masses on heavy quark mass are plotted
in FIG.~\ref{DepMcG0Sample1N}. We use $H$ to represent the heavy
pseudo-scalar meson and $H^*$ the heavy vector meson. One can see that
when the quark mass tends to infinity, the mass splitting between $H$
and $H^*$ meson vanishes.
On the other hand if we use the interaction in Eq.~(\ref{L_Qq0}) and
keep the parameter $G_2$ unchanged vs the quark masses, i.e.
\[
  G_2=G_V+\frac{2h}{m_u^2}=5.41/\Lambda^2,
\]
we observe a mass crossing of the $H$ meson with $H^*$ meson mass as
the heavy quark mass increase.  Beyond the crossing point, the mass
relation is reverted with the $H$ meson above the $H^*$. the mass
curve of $H$ will further reach the mass threshold and no $H$ bound
state exists beyond. So, a naive generalization of the NJL interaction
from light quark sector to the heavy is inappropriate.

As a comparison, we also checked with the interaction of a mass
independent vector interaction in heavy meson sector,
i.e. Eq.~(\ref{L_Qq0}) with $G_1=G_2=g_3\Lambda^{-2}$.  The
interaction in light meson sector is Eq.~(\ref{L_basic}) with a
different set of couplings $G_1=g_1\Lambda^{-2}$ and
$G_2=g_2\Lambda^{-2}$.  The parameters are
\begin{equation}
  \begin{aligned}
    m_{u/d}^0=& 3.36\text{MeV}, & m_s^0=& 81.7\text{MeV},\\
    m_c &= 1.68\text{GeV}, & m_b=& 5.00\text{GeV}, & \Lambda=&0.7\text{GeV},\\
    g_1 &= 2.52, & g_2=&5.82 & g_3=&2.53.
  \end{aligned}
\end{equation}
The result is shown in the cal.~II column in
Tab.~\ref{TabResultsG0}. We notice that the mass splitting between the
heavy pseudo-scalar meson $D$ (or $B$) and its vector partner $D^*$
(or $B^*$) differs from the empirical data by roughly a factor
$1.5$. For the $D$ and $D^*$, it is $210$MeV comparing to the
empirical data $150$MeV, and for the $B$ and $B^*$, it is $80$MeV
comparing to $50$MeV. To reduce the mass splitting, one may decrease
the coupling $g_V$. But $g_V$ can not be too small, otherwise the interaction
will not be strong enough to bound the $D^*$ meson.

\section{Conclusion}

In this work, we have studied light and heavy mesons in a unified
frame with the NJL model. We have followed a traditional approach of
solving the DSE and BSE. We have used a 3 dimensional cutoff to
adequately regularize the integrals when heavy quarks are involved.

We have investigated the heavy quark spin symmetry in the heavy quark
limit.  We find that, in the heavy quark limit, the pseudo-scalar
meson and its vector partner will have an identical mass equation only
if the NJL interaction is a color-octet vector interaction which can
be recognized as a approximation of a single-gluon exchange
interaction.

Then we propose an extension to the NJL interaction as in
Eq.~(\ref{L_Qq0}) which introduces the $1/m_q$ correction to the quark
current. The mass dependence suppresses the axial-vector current
interaction to guarantee that the heavy quark spin symmetry still
holds in the heavy quark limit.

We have performed numerical calculations to the light and heavy
pseudo-scalar and vector meson, both for their masses and the weak
decay constants. The mass spectra fit the experimental data quite
well.  But the weak decay constants always show a large discrepancy to
the experiments. A possible explanation is that the momentum cutoff in
heavy sector is larger than in light sector, which reflects the fact
that the size of a heavy meson is relatively small. The issue can be
studied using some more realistic interaction other than the contact
one.

\begin{appendix}
\section{The Current Condensates in 3D Cutoff}

In BSE, we need to calculate the loop integral
\begin{equation}
  J(\Gamma,\Gamma',m,m')=2iN_C\mathrm{tr}\int \frac{\mathrm{d^4}p}{(2\pi)^4}
  \left[\Gamma\frac{1}{(\slashed{p}+\frac{1}{2}\slashed{q})-m+i\epsilon}
    \Gamma'\frac{1}{(\slashed{p}-\frac{1}{2}\slashed{q})-m'+i\epsilon}
  \right],
\end{equation}
where $\Gamma$ and $\Gamma'$ are the interaction vertices.  For
pseudo-scalar mesons, we have \cite{Klimt:1989pm}
\begin{align}
  J_{PP}&=\frac{1}{2}[I_1(m)+I_1(m')]+[(m-m')^2-q^2]I_2(m,m',q^2),\\
  J_{PA,\mu}&=q_\mu(m+m')\left[1-\frac{(m-m')^2}{q^2}\right]I_2(q^2,m,m')
  +q_\mu\frac{m-m'}{2q^2}(I_1(m)-I_1(m')),\\
  J_{AA}^{L}&=\frac{(m^2-m'^2)^2}{q^2}(I_2-I_2^0)-(m+m')^2I_2 .
\end{align}
For vector mesons, the loop integral is healed by subtracting a
certain term $J_{VV}^T\rightarrow
J_{VV}^T-J_{VV}^{(T)}(q=0)+J_{VV}^{(L)}(q=0)$ and one can obtain,
\begin{equation}
  J_{VV}^{T}
  =\frac{1}{3}\left[2(m^2+m'^2)(I_2-I_2^0)-[3(m-m')^2-2q^2]I_2
    -\frac{(m^2-m'^2)^2}{q^2}(I_2-I_2^0)+4(m^2-m'^2)^2{I_2^0}'\right].
\end{equation}
The subtracted term tends to zero when one quark mass tends to
infinity.  The integrations involved are,
\begin{align}
&I_1(m)=8iN_C\int\frac{\mathrm{d}^4p}{(2\pi)^4}\frac{1}{(p^2-m^2+i\epsilon)}\\
&I_2(m,m',q^2)=4iN_C\int\frac{\mathrm{d}^4p}{(2\pi)^4}\frac{1}{[(p+\frac{1}{2}q)^2-m^2+i\epsilon][(p-\frac{1}{2}q)^2-m'^2+i\epsilon]}
\end{align}
and define, $I_2^0(m,m')\equiv I_2(m,m',0)$ and ${I_2^0}'=\mathrm{d}I_2/\mathrm{d}q^2|_{q^2=0}$.
After a calculation, one can find, when  $(m'-m)^2<q^2<(m+m')^2$,
\begin{align}
&I_1(m)=\frac{N_C}{4\pi^2}\int_{4m^2}^{4(\Lambda^2+m^2)}\sqrt{1-\frac{4m^2}{\kappa^2}}\mathrm{d}\kappa^2\\
&I_2(m,m',q^2)=-\frac{N_C}{4\pi^2} \int_{(m+m')^2}^{(\sqrt{\Lambda^2+m^2}+\sqrt{\Lambda^2+m'^2})^2}\frac{\sqrt{1-2\frac{m^2+m'^2}{\kappa^2}+\big(\frac{m^2-m'^2}{\kappa^2}\big)^2}}{\kappa^2-q^2 }\mathrm{d}\kappa^2
\end{align}
in which the $\Lambda^2$ is the 3 dimensional cutoff. The same expression can be applied to the case $q^2<(m-m')^2$.

The integration involved in $0^-$, $1^-$ sector is,
\begin{equation}
\begin{split}
j(l_0)&=4iN_C\int\frac{\mathrm{d}^4k}{(2\pi)^4}\frac{k\cdot v-m_q}{(k^2-m_q^2+i\epsilon)(v\cdot k +l_0+i\epsilon)}
\end{split}
\end{equation}
Assuming $v=(1,\vec{0})$, and integrating out $k_0$ below the threshold $l_0<m_q$, one can find,
\begin{equation}
j(l_0)=\frac{4N_C}{(2\pi)^4}\pi\int\mathrm{d}^3k\frac{\sqrt{\mathbf{k}^2+m^2}+m_q}{\sqrt{\mathbf{k}^2+m_q^2}(\sqrt{\mathbf{k}^2+m_q^2}-l_0)}
\end{equation}
Introducing the 3D cutoff, one can get,
\begin{equation}
j(l_0)=\frac{N_C}{(2\pi)^2}\int_{4m_q^2}^{4(\Lambda^2+m_q^2)}\frac{\kappa+2m_q}{2\kappa-4l_0}\sqrt{1-\frac{4m^2}{\kappa}}\mathrm{d}\kappa^2
\end{equation}
In which,
\begin{equation}
\kappa^2=4\mathbf{k}^2+4m_q^2
\end{equation}

\end{appendix}

\bibliography{hnjl}

\begin{thebibliography}{33}
\expandafter\ifx\csname natexlab\endcsname\relax\def\natexlab#1{#1}\fi
\expandafter\ifx\csname bibnamefont\endcsname\relax
  \def\bibnamefont#1{#1}\fi
\expandafter\ifx\csname bibfnamefont\endcsname\relax
  \def\bibfnamefont#1{#1}\fi
\expandafter\ifx\csname citenamefont\endcsname\relax
  \def\citenamefont#1{#1}\fi
\expandafter\ifx\csname url\endcsname\relax
  \def\url#1{\texttt{#1}}\fi
\expandafter\ifx\csname urlprefix\endcsname\relax\def\urlprefix{URL }\fi
\providecommand{\bibinfo}[2]{#2}
\providecommand{\eprint}[2][]{\url{#2}}

\bibitem[{\citenamefont{Tornqvist}(2004)}]{Tornqvist:2004qy}
\bibinfo{author}{\bibfnamefont{N.~A.} \bibnamefont{Tornqvist}},
  \bibinfo{journal}{Phys. Lett.} \textbf{\bibinfo{volume}{B590}},
  \bibinfo{pages}{209} (\bibinfo{year}{2004}), \eprint{hep-ph/0402237}.

\bibitem[{\citenamefont{Swanson}(2004)}]{Swanson:2003tb}
\bibinfo{author}{\bibfnamefont{E.~S.} \bibnamefont{Swanson}},
  \bibinfo{journal}{Phys. Lett.} \textbf{\bibinfo{volume}{B588}},
  \bibinfo{pages}{189} (\bibinfo{year}{2004}), \eprint{hep-ph/0311229}.

\bibitem[{\citenamefont{Liu et~al.}(2009)\citenamefont{Liu, Luo, Liu, and
  Zhu}}]{Liu:2008tn}
\bibinfo{author}{\bibfnamefont{X.}~\bibnamefont{Liu}},
  \bibinfo{author}{\bibfnamefont{Z.-G.} \bibnamefont{Luo}},
  \bibinfo{author}{\bibfnamefont{Y.-R.} \bibnamefont{Liu}}, \bibnamefont{and}
  \bibinfo{author}{\bibfnamefont{S.-L.} \bibnamefont{Zhu}},
  \bibinfo{journal}{Eur. Phys. J.} \textbf{\bibinfo{volume}{C61}},
  \bibinfo{pages}{411} (\bibinfo{year}{2009}), \eprint{0808.0073}.

\bibitem[{\citenamefont{Liu and Zhang}(2009)}]{Liu:2008qb}
\bibinfo{author}{\bibfnamefont{Y.-R.} \bibnamefont{Liu}} \bibnamefont{and}
  \bibinfo{author}{\bibfnamefont{Z.-Y.} \bibnamefont{Zhang}},
  \bibinfo{journal}{Phys. Rev.} \textbf{\bibinfo{volume}{C79}},
  \bibinfo{pages}{035206} (\bibinfo{year}{2009}), \eprint{0805.1616}.

\bibitem[{\citenamefont{Yu et~al.}(2012)\citenamefont{Yu, Wang, Chen, and
  Deng}}]{Yu:2011wb}
\bibinfo{author}{\bibfnamefont{S.-H.} \bibnamefont{Yu}},
  \bibinfo{author}{\bibfnamefont{B.-K.} \bibnamefont{Wang}},
  \bibinfo{author}{\bibfnamefont{X.-L.} \bibnamefont{Chen}}, \bibnamefont{and}
  \bibinfo{author}{\bibfnamefont{W.-Z.} \bibnamefont{Deng}},
  \bibinfo{journal}{Chin. Phys.} \textbf{\bibinfo{volume}{C36}},
  \bibinfo{pages}{25} (\bibinfo{year}{2012}), \eprint{1104.4535}.

\bibitem[{\citenamefont{Nambu and
  Jona-Lasinio}(1961{\natexlab{a}})}]{Nambu:1961tp}
\bibinfo{author}{\bibfnamefont{Y.}~\bibnamefont{Nambu}} \bibnamefont{and}
  \bibinfo{author}{\bibfnamefont{G.}~\bibnamefont{Jona-Lasinio}},
  \bibinfo{journal}{Phys. Rev.} \textbf{\bibinfo{volume}{122}},
  \bibinfo{pages}{345} (\bibinfo{year}{1961}{\natexlab{a}}).

\bibitem[{\citenamefont{Nambu and
  Jona-Lasinio}(1961{\natexlab{b}})}]{Nambu:1961fr}
\bibinfo{author}{\bibfnamefont{Y.}~\bibnamefont{Nambu}} \bibnamefont{and}
  \bibinfo{author}{\bibfnamefont{G.}~\bibnamefont{Jona-Lasinio}},
  \bibinfo{journal}{Phys. Rev.} \textbf{\bibinfo{volume}{124}},
  \bibinfo{pages}{246} (\bibinfo{year}{1961}{\natexlab{b}}).

\bibitem[{\citenamefont{Klevansky}(1992)}]{Klevansky:1992qe}
\bibinfo{author}{\bibfnamefont{S.~P.} \bibnamefont{Klevansky}},
  \bibinfo{journal}{Rev. Mod. Phys.} \textbf{\bibinfo{volume}{64}},
  \bibinfo{pages}{649} (\bibinfo{year}{1992}).

\bibitem[{\citenamefont{Hatsuda and Kunihiro}(1994)}]{Hatsuda:1994pi}
\bibinfo{author}{\bibfnamefont{T.}~\bibnamefont{Hatsuda}} \bibnamefont{and}
  \bibinfo{author}{\bibfnamefont{T.}~\bibnamefont{Kunihiro}},
  \bibinfo{journal}{Phys. Rept.} \textbf{\bibinfo{volume}{247}},
  \bibinfo{pages}{221} (\bibinfo{year}{1994}), \eprint{hep-ph/9401310}.

\bibitem[{\citenamefont{Vogl and Weise}(1991)}]{Vogl:1991qt}
\bibinfo{author}{\bibfnamefont{U.}~\bibnamefont{Vogl}} \bibnamefont{and}
  \bibinfo{author}{\bibfnamefont{W.}~\bibnamefont{Weise}},
  \bibinfo{journal}{Prog. Part. Nucl. Phys.} \textbf{\bibinfo{volume}{27}},
  \bibinfo{pages}{195} (\bibinfo{year}{1991}).

\bibitem[{\citenamefont{Bernard and Meissner}(1988)}]{Bernard:1988db}
\bibinfo{author}{\bibfnamefont{V.}~\bibnamefont{Bernard}} \bibnamefont{and}
  \bibinfo{author}{\bibfnamefont{U.~G.} \bibnamefont{Meissner}},
  \bibinfo{journal}{Nucl. Phys.} \textbf{\bibinfo{volume}{A489}},
  \bibinfo{pages}{647} (\bibinfo{year}{1988}).

\bibitem[{\citenamefont{Blin et~al.}(1990)\citenamefont{Blin, Hiller, and
  da~Providencia}}]{Blin:1990um}
\bibinfo{author}{\bibfnamefont{A.~H.} \bibnamefont{Blin}},
  \bibinfo{author}{\bibfnamefont{B.}~\bibnamefont{Hiller}}, \bibnamefont{and}
  \bibinfo{author}{\bibfnamefont{J.}~\bibnamefont{da~Providencia}},
  \bibinfo{journal}{Phys. Lett.} \textbf{\bibinfo{volume}{B241}},
  \bibinfo{pages}{1} (\bibinfo{year}{1990}).

\bibitem[{\citenamefont{Takizawa et~al.}(1991)\citenamefont{Takizawa, Kubodera,
  and Myhrer}}]{Takizawa:1991mx}
\bibinfo{author}{\bibfnamefont{M.}~\bibnamefont{Takizawa}},
  \bibinfo{author}{\bibfnamefont{K.}~\bibnamefont{Kubodera}}, \bibnamefont{and}
  \bibinfo{author}{\bibfnamefont{F.}~\bibnamefont{Myhrer}},
  \bibinfo{journal}{Phys. Lett.} \textbf{\bibinfo{volume}{B261}},
  \bibinfo{pages}{221} (\bibinfo{year}{1991}).

\bibitem[{\citenamefont{Bernard et~al.}(1988)\citenamefont{Bernard, Jaffe, and
  Meissner}}]{Bernard:1987sg}
\bibinfo{author}{\bibfnamefont{V.}~\bibnamefont{Bernard}},
  \bibinfo{author}{\bibfnamefont{R.~L.} \bibnamefont{Jaffe}}, \bibnamefont{and}
  \bibinfo{author}{\bibfnamefont{U.~G.} \bibnamefont{Meissner}},
  \bibinfo{journal}{Nucl. Phys.} \textbf{\bibinfo{volume}{B308}},
  \bibinfo{pages}{753} (\bibinfo{year}{1988}).

\bibitem[{\citenamefont{Klimt et~al.}(1990)\citenamefont{Klimt, Lutz, Vogl, and
  Weise}}]{Klimt:1989pm}
\bibinfo{author}{\bibfnamefont{S.}~\bibnamefont{Klimt}},
  \bibinfo{author}{\bibfnamefont{M.~F.~M.} \bibnamefont{Lutz}},
  \bibinfo{author}{\bibfnamefont{U.}~\bibnamefont{Vogl}}, \bibnamefont{and}
  \bibinfo{author}{\bibfnamefont{W.}~\bibnamefont{Weise}},
  \bibinfo{journal}{Nucl. Phys.} \textbf{\bibinfo{volume}{A516}},
  \bibinfo{pages}{429} (\bibinfo{year}{1990}).

\bibitem[{\citenamefont{Eguchi and Sugawara}(1974)}]{Eguchi:1974cg}
\bibinfo{author}{\bibfnamefont{T.}~\bibnamefont{Eguchi}} \bibnamefont{and}
  \bibinfo{author}{\bibfnamefont{H.}~\bibnamefont{Sugawara}},
  \bibinfo{journal}{Phys. Rev.} \textbf{\bibinfo{volume}{D10}},
  \bibinfo{pages}{4257} (\bibinfo{year}{1974}).

\bibitem[{\citenamefont{Ebert and Volkov}(1983)}]{Ebert:1982pk}
\bibinfo{author}{\bibfnamefont{D.}~\bibnamefont{Ebert}} \bibnamefont{and}
  \bibinfo{author}{\bibfnamefont{M.~K.} \bibnamefont{Volkov}},
  \bibinfo{journal}{Z. Phys.} \textbf{\bibinfo{volume}{C16}},
  \bibinfo{pages}{205} (\bibinfo{year}{1983}).

\bibitem[{\citenamefont{Ebert and Reinhardt}(1986)}]{Ebert:1985kz}
\bibinfo{author}{\bibfnamefont{D.}~\bibnamefont{Ebert}} \bibnamefont{and}
  \bibinfo{author}{\bibfnamefont{H.}~\bibnamefont{Reinhardt}},
  \bibinfo{journal}{Nucl. Phys.} \textbf{\bibinfo{volume}{B271}},
  \bibinfo{pages}{188} (\bibinfo{year}{1986}).

\bibitem[{\citenamefont{Reinhardt and Alkofer}(1988)}]{Reinhardt:1988xu}
\bibinfo{author}{\bibfnamefont{H.}~\bibnamefont{Reinhardt}} \bibnamefont{and}
  \bibinfo{author}{\bibfnamefont{R.}~\bibnamefont{Alkofer}},
  \bibinfo{journal}{Phys. Lett.} \textbf{\bibinfo{volume}{B207}},
  \bibinfo{pages}{482} (\bibinfo{year}{1988}).

\bibitem[{\citenamefont{Bijnens}(1996)}]{Bijnens:1995ww}
\bibinfo{author}{\bibfnamefont{J.}~\bibnamefont{Bijnens}},
  \bibinfo{journal}{Phys. Rept.} \textbf{\bibinfo{volume}{265}},
  \bibinfo{pages}{369} (\bibinfo{year}{1996}), \eprint{hep-ph/9502335}.

\bibitem[{\citenamefont{Isgur and Wise}(1989)}]{Isgur:1989vq}
\bibinfo{author}{\bibfnamefont{N.}~\bibnamefont{Isgur}} \bibnamefont{and}
  \bibinfo{author}{\bibfnamefont{M.~B.} \bibnamefont{Wise}},
  \bibinfo{journal}{Phys. Lett.} \textbf{\bibinfo{volume}{B232}},
  \bibinfo{pages}{113} (\bibinfo{year}{1989}).

\bibitem[{\citenamefont{Manohar and Wise}(2000)}]{manoh}
\bibinfo{author}{\bibfnamefont{A.}~\bibnamefont{Manohar}} \bibnamefont{and}
  \bibinfo{author}{\bibfnamefont{M.}~\bibnamefont{Wise}},
  \emph{\bibinfo{title}{Heavy Quark Physics}} (\bibinfo{publisher}{Cambrige
  University Press}, \bibinfo{year}{2000}).

\bibitem[{\citenamefont{Neubert}(1994)}]{Neubert:1993mb}
\bibinfo{author}{\bibfnamefont{M.}~\bibnamefont{Neubert}},
  \bibinfo{journal}{Phys. Rept.} \textbf{\bibinfo{volume}{245}},
  \bibinfo{pages}{259} (\bibinfo{year}{1994}), \eprint{hep-ph/9306320}.

\bibitem[{\citenamefont{Ebert et~al.}(1995)\citenamefont{Ebert, Feldmann,
  Friedrich, and Reinhardt}}]{Ebert:1994tv}
\bibinfo{author}{\bibfnamefont{D.}~\bibnamefont{Ebert}},
  \bibinfo{author}{\bibfnamefont{T.}~\bibnamefont{Feldmann}},
  \bibinfo{author}{\bibfnamefont{R.}~\bibnamefont{Friedrich}},
  \bibnamefont{and}
  \bibinfo{author}{\bibfnamefont{H.}~\bibnamefont{Reinhardt}},
  \bibinfo{journal}{Nucl. Phys.} \textbf{\bibinfo{volume}{B434}},
  \bibinfo{pages}{619} (\bibinfo{year}{1995}), \eprint{hep-ph/9406220}.

\bibitem[{\citenamefont{Mota and Arriola}(2007)}]{Mota:2006ex}
\bibinfo{author}{\bibfnamefont{A.~L.} \bibnamefont{Mota}} \bibnamefont{and}
  \bibinfo{author}{\bibfnamefont{E.~R.} \bibnamefont{Arriola}},
  \bibinfo{journal}{Eur. Phys. J.} \textbf{\bibinfo{volume}{A31}},
  \bibinfo{pages}{711} (\bibinfo{year}{2007}), \eprint{hep-ph/0610146}.

\bibitem[{\citenamefont{Ivanov et~al.}(1998)\citenamefont{Ivanov, Kalinovsky,
  Maris, and Roberts}}]{Ivanov:1997yg}
\bibinfo{author}{\bibfnamefont{M.~A.} \bibnamefont{Ivanov}},
  \bibinfo{author}{\bibfnamefont{Y.~L.} \bibnamefont{Kalinovsky}},
  \bibinfo{author}{\bibfnamefont{P.}~\bibnamefont{Maris}}, \bibnamefont{and}
  \bibinfo{author}{\bibfnamefont{C.~D.} \bibnamefont{Roberts}},
  \bibinfo{journal}{Phys. Lett.} \textbf{\bibinfo{volume}{B416}},
  \bibinfo{pages}{29} (\bibinfo{year}{1998}), \eprint{nucl-th/9704039}.

\bibitem[{\citenamefont{Ivanov et~al.}(1999)\citenamefont{Ivanov, Kalinovsky,
  and Roberts}}]{Ivanov:1998ms}
\bibinfo{author}{\bibfnamefont{M.~A.} \bibnamefont{Ivanov}},
  \bibinfo{author}{\bibfnamefont{Y.~L.} \bibnamefont{Kalinovsky}},
  \bibnamefont{and} \bibinfo{author}{\bibfnamefont{C.~D.}
  \bibnamefont{Roberts}}, \bibinfo{journal}{Phys. Rev.}
  \textbf{\bibinfo{volume}{D60}}, \bibinfo{pages}{034018}
  (\bibinfo{year}{1999}), \eprint{nucl-th/9812063}.

\bibitem[{\citenamefont{Roberts and Schmidt}(2000)}]{Roberts:2000aa}
\bibinfo{author}{\bibfnamefont{C.~D.} \bibnamefont{Roberts}} \bibnamefont{and}
  \bibinfo{author}{\bibfnamefont{S.~M.} \bibnamefont{Schmidt}},
  \bibinfo{journal}{Prog. Part. Nucl. Phys.} \textbf{\bibinfo{volume}{45}},
  \bibinfo{pages}{S1} (\bibinfo{year}{2000}), \eprint{nucl-th/0005064}.

\bibitem[{\citenamefont{Nguyen et~al.}(2009)\citenamefont{Nguyen, Souchlas, and
  Tandy}}]{Nguyen:2009if}
\bibinfo{author}{\bibfnamefont{T.}~\bibnamefont{Nguyen}},
  \bibinfo{author}{\bibfnamefont{N.~A.} \bibnamefont{Souchlas}},
  \bibnamefont{and} \bibinfo{author}{\bibfnamefont{P.~C.} \bibnamefont{Tandy}},
  \bibinfo{journal}{AIP Conf. Proc.} \textbf{\bibinfo{volume}{1116}},
  \bibinfo{pages}{327} (\bibinfo{year}{2009}), \eprint{0904.3345}.

\bibitem[{\citenamefont{Nakamura et~al.}(2010)}]{Nakamura:2010zzi}
\bibinfo{author}{\bibfnamefont{K.}~\bibnamefont{Nakamura}} \bibnamefont{et~al.}
  (\bibinfo{collaboration}{Particle Data Group}), \bibinfo{journal}{J. Phys.}
  \textbf{\bibinfo{volume}{G37}}, \bibinfo{pages}{075021}
  (\bibinfo{year}{2010}).

\bibitem[{\citenamefont{Gamiz et~al.}(2009)\citenamefont{Gamiz, Davies, Lepage,
  Shigemitsu, and Wingate}}]{Gamiz:2009ku}
\bibinfo{author}{\bibfnamefont{E.}~\bibnamefont{Gamiz}},
  \bibinfo{author}{\bibfnamefont{C.~T.~H.} \bibnamefont{Davies}},
  \bibinfo{author}{\bibfnamefont{G.~P.} \bibnamefont{Lepage}},
  \bibinfo{author}{\bibfnamefont{J.}~\bibnamefont{Shigemitsu}},
  \bibnamefont{and} \bibinfo{author}{\bibfnamefont{M.}~\bibnamefont{Wingate}}
  (\bibinfo{collaboration}{HPQCD}), \bibinfo{journal}{Phys. Rev.}
  \textbf{\bibinfo{volume}{D80}}, \bibinfo{pages}{014503}
  (\bibinfo{year}{2009}), \eprint{0902.1815}.

\bibitem[{\citenamefont{Bazavov et~al.}(2011)}]{Bazavov:2011aa}
\bibinfo{author}{\bibfnamefont{A.}~\bibnamefont{Bazavov}} \bibnamefont{et~al.}
  (\bibinfo{collaboration}{Fermilab Lattice and MILC}) (\bibinfo{year}{2011}),
  \eprint{1112.3051}.

\bibitem[{\citenamefont{Na et~al.}(2012)}]{Na:2012kp}
\bibinfo{author}{\bibfnamefont{H.}~\bibnamefont{Na}} \bibnamefont{et~al.}
  (\bibinfo{year}{2012}), \eprint{1202.4914}.

\end{thebibliography}

\end{document}